\documentclass[12pt]{article}

\usepackage{sectsty}
\usepackage{fullpage}
\usepackage[export]{adjustbox}
\usepackage{graphicx}

\usepackage{cancel,enumerate}
\usepackage[affil-it]{authblk}

\usepackage{amsmath}
\usepackage{amstext, amssymb}
\usepackage{eufrak}
\usepackage[mathscr]{euscript}

\usepackage{tikz}[2013/12/13]
\usetikzlibrary{calc}
\usetikzlibrary{matrix}
\definecolor{dark}{RGB}{0, 133, 202 }
\definecolor{light}{RGB}{112, 155, 230}

\newcommand{\xx}{\mathbf{x}}
\newcommand{\FF}{\mathbf{F}}
\newcommand{\bnab}{\mathbf{\nabla}}
\newcommand{\oE}{\overline{E}}
\newcommand{\dd}{{\rm d}}

\begin{document}

\title{Molecular geometry and vibrational frequencies by parallel sampling}
\author[1]{Jan Vrbik}
\affil[1]{Department of Mathematics and Statistics\\ Brock University, Canada}
\date{\today}
\maketitle

\begin{abstract}
Quantum Monte Carlo is an efficient technique for finding the ground-state
energy and related properties of small molecules.  A major challenge remains
in accurate determination of a molecule's geometry, i.e.  the optimal
location of its individual nuclei and the frequencies of their vibration.
The aim of this article is to describe a simple technique to accurately
establish such properties.  This is achieved by varying the trial function to
accommodate changing geometry, thereby removing a source of rather
unpleasant singularities which arise when the trial function is fixed (the
traditional approach).\medskip

\noindent\textbf{Keywords:} QMC; molecular geometry; vibrational frequencies; energy
derivatives
\end{abstract}

\section{Introduction}

In this article we describe a new technique of\textsc{\ parallel sampling } for computing various properties (those related to location and vibration of nuclei in particular) of small molecules.  But first we present (for the benefit of a non-specialist) a brief review of general Quantom Monte Carlo.

To find the \textsc{ground-state energy} of a molecule given its \textsc{geometry} (i.e.  distances between its individual nuclei, for this purpose considered fixed -- this constitutes the so-called \textsc{clamped}
approximation), one has to solve the following eigenvalue problem
\begin{equation}
-\frac{1}{2}\bnab^{2}\Phi (\xx)+V(\xx)\Phi(\xx)=\mathcal{E}\,\Phi(\xx)  \label{schr}
\end{equation}
where $\Phi(\xx)$ is an (unknown) function of the locations of the molecule's (say $n$) \emph{electrons} collectively denoted $\xx$$,\bnab^{2}$ is the corresponding$3n$-dimensional Laplace operator, $V(\xx)$ is the molecule's electro-static potential and $\mathcal{E}$ is the corresponding \emph{lowest} eigenvalue.  The trouble is that this equation can be solved only approximately (usually using the \textsc{variational principle}) by a function we denote $\Psi(\xx)$ and call the \textsc{trial function}.  In this approximate solution, $\mathcal{E}$ itself becomes a function of $\xx$ (the so-called local energy, defined shortly) which is only approximately `constant'. We can improve the accuracy of such a solution by increasing the complexity of $\Psi(\xx)$, correspondingly increasing the challenge of optimizing its parameters.

Having such variational solution (of whatever quality), Diffusion Monte Carlo aims to substantially improve the corresponding estimates of $\mathcal{E}$ (and eventually of its derivatives) by the following procedure (see, for example \cite{anderson} and \cite{kalos}):

\begin{enumerate}
\item Based on $\Psi(\xx)$, it defines two more functions, namely
\begin{enumerate}
\item the so-called \textsc{drift} function%
\begin{equation}
\FF(\xx):=\frac{\bnab~\Psi(\xx)}{\Psi(\xx)}
\end{equation}
where $\bnab$ represents the vector of all partial derivatives (with respect to each component of $\xx$) -- note that this makes $\FF(\xx)$ a \emph{vector} with $3n$ components
\item and the \textsc{local energy}
\begin{equation}
E(\xx):=-\frac{1}{2}\frac{\bnab^{2}\Psi(\xx)}{\Psi(\xx)}+V(\xx)
\end{equation}
which is a scalar function of$\xx$.  We mention in passing that 
\begin{equation}
\frac{\mathop E(\xx)\Psi(\xx)^{2}d\xx}{\mathop
\Psi(\xx)^{2}d\xx}  \label{variat}
\end{equation}%
represents the so-called \textsc{variational} approximation to $\mathcal{E}$ (always higher than the correct value), where the integration is over all space.
\end{enumerate}
\item Starting with a rather arbitrarily chosen collection (called an \textsc{ensemble}, which should be as large as our computer can handle) of \textsc{configurations} (numerical values of $\xx$), and using a small value of $\tau$ (called \textsc{time step}), we then advance each configuration $\xx$ to a new $\xx^{\prime }$ by
\begin{equation}
\xx^{\prime }=\xx+\tau ~\FF(\xx)+\mathbf{\chi }
\label{move}
\end{equation}
where $\mathbf{\chi }$ is a \emph{random} vector with$3n$ components, drawn independently from the Normal distribution with the mean of $0$ and the standard deviation of $\sqrt{\tau }$. The $\tau ~\FF(\xx)$ part of this `move' is the actual `drift', while the $\mathbf{\chi }$ part is referred to as \textsc{diffusion}.  We then compute, for each resulting new configuration, the value of $E(\xx^{\prime });$ this constitutes a so-called \textsc{iteration}.
\item We then repeat as many of these iterations as possible ($\xx^{\prime}$ always becomes the `old' $\xx$ of the next iteration).  For each configuration, we also keep track of the sum of all (current and past) values of the local energy, \emph{depreciated} by the factor of
\begin{equation}
1-\frac{a~\tau}{1-\ln \tau}  \label{Efac}
\end{equation}
where $a$ is an empirically chosen constant.  To be more specific: in each iteration the old sum is updated by multiplying it by \eqref{Efac} and adding to it the new value of $E(\xx^{\prime })$; the result will be called $S(\xx^{\prime })$. We then compute the iteration's \emph{weighted} average of the current $E(\xx^{\prime })$ values by
\begin{equation}
\oE:=\frac{\sum E(\xx^{\prime })\exp (-\tau~S(\xx^{\prime }))}{\sum \exp (-\tau ~S(\xx^{\prime }))}
\label{aver}
\end{equation}
where the summation is over all configurations of the ensemble and 
\begin{equation}
W(\xx^{\prime }):=\exp (-\tau ~S(\xx^{\prime }))  \label{weights}
\end{equation}
are the corresponding weights.  Note that the simple average of the $E(\xx^{\prime })$ values would estimate (\ref{variat}) instead; this is true in general: whenever we replace $W(\xx^{\prime })$ by $1$ (which corresponds to \emph{regular} averaging) we get the so-called (and rather inferior) \emph{variational} estimates -- this applies to our subsequent estimation of various derivatives of $\mathcal{E}$ (see \cite{vrbik}) as well.
\item By monitoring the values of $\oE$ from iteration to iteration, we can easily notice that after a while the process stabilizes (the values of $\oE$ no longer systematically increase or decrease; they only randomly fluctuate around some mean value).  At the end of the simulation (after thousands of iterations) we again average (this time in the regular manner) the iteration averages themselves (but only the \textsc{equilibrated} ones, after stabilization), denoting the resulting \textsc{grand mean} by $\overline{\oE}(\tau )$.
\item Even this $\overline{\oE}(\tau)$ is not yet our final estimate of $\mathcal{E}$, since it contains a $\tau$-related \textsc{bias}.  To remove it, we have to repeat the simulation with several \emph{different} values of $\tau$ (three to five such values are usually sufficient) and extrapolate the corresponding grand means to $\tau=0$ by a simple (usually quadratic) regression.
\end{enumerate}

We should mention (since it is far from obvious) that this corresponds to replacing, in (\ref{variat}), $\Psi(\xx)^{2}$ by the so-called \textsc{mixed} distribution, proportional to the \emph{product} of the exact solution $\Phi(\xx)$ and the variational solution $\Psi(\xx)$. Since $\Phi(\xx)$ is thus forced to have the same nodes as $\Psi(\xx)$, this introduces a small (in our case almost negligible) \textsc{fixed-node} error into the final estimate of $\mathcal{E}$. Subsequently (and due to this error somehow incorrectly), we call the corresponding results \emph{exact}, those using regular averaging in (\ref{aver}) are called \emph{variational}.

Note that the proposed technique has no need for the so-called \textsc{pure} sampling from a distribution proportional to $\Phi(\xx)^{2};$ such sampling does \emph{not} properly estimate $\mathcal{E}$ and is useful only in the context of Hellmann-Feynman Theorem.

\section{Parallel simulation}

We now consider the case of $V(\xx)$, and consequently of $\Phi(\xx)$ and $\mathcal{E}$, being functions of a parameter $\lambda$ (the technique can be easily extended to deal with \emph{several} such parameters).  Usually $\lambda$ represents the strength of an external perturbation (such as a uniform electrical field), but it can also be the distance between two of the molecule's nuclei (as mentioned already, these distances constitute what we call the molecule's geometry).  We assume that the trial solution $\Psi(\xx)$ is \emph{also} a function of $\lambda$ (e.g.  atomic orbitals of the trial function are centered on individual nuclei, and move with these).  This implies that both $\FF(\xx)$ and $E(\xx)$ also vary with $\lambda$ (now, an implicit argument of both functions).

The objective of this article is to find a good approximation to the first, second and perhaps higher \emph{derivatives} of the ground-state energy with respect to $\lambda$ (at $\lambda=0$ in the case of external perturbations, or at $\lambda =\lambda_{0}$, where $\lambda_{0}$ represents the initial inter-nuclear distance).  We want to emphasize that such derivatives yield \emph{most} of the basic molecular properties, e.g. the dipole moment, polarizability, harmonic and unharmonic frequencies etc., including the molecule's \emph{optimal} geometry.

We achieve this goal by a modification of the so-called \textsc{correlated sampling} of \cite{umri} and \cite{fili}; instead of using small but finite separation between different values of $\lambda$ (as they have done), we propose to make these differences truly `infinitesimal' (which is the main thrust of this article).  This alleviates the problem of sudden divergence of parallel configurations (as discussed in \cite{assa1} and \cite{assa2}), and eliminates the need not only for subsequent finite-step differentiation but, most importantly, for the so-called space-wrapping of electrons\ (see \cite{umri}).  We also deliberately avoid employing the Hellmann-Feynman theorem (recommended only for high-quality trial functions, e.g.  those of \cite{bad1} and \cite{bad2}), since its application may seriously bias our results (see 
\cite{bad0}).

We now describe a procedure to accurately estimate the ground-state energy and its \emph{first} and \emph{second} derivatives with respect to $\lambda$; extending this to the case of higher derivatives is then fairly routine.\medskip

To accomplish the former, one has to perform the five steps of the previous section, with the following extensions:

\begin{enumerate}
\item Throughout the simulation, each configuration $\xx$ splits into three (\emph{initially} identical `companions'), which are then advanced `in parallel' using thee \emph{slightly} different values of $\lambda$ (namely $\lambda_{0}-\varepsilon $, $\lambda_{0}$, and $\lambda _{0}+\varepsilon$) in the drift part of (\ref{move}), but the \emph{same} random vector $\mathbf{\chi }$ in the diffusion part.  The value of $\varepsilon$ (aiming for \emph{practically} `infinitesimal') is chosen to decrease with $\tau$ in the following manner
\begin{equation}
\varepsilon = b\sqrt{\tau }  \label{epsi}
\end{equation}
where $b$ is a suitable constant (to keep$\varepsilon$ in the $10^{-5}$ to $10^{-4}$ range).
\item Similarly, the$E(\xx)$ and$S(\xx)$ are each computed in the corresponding triplicate (referring to them as $E_{-}$, $E_{0}$, and $E_{+}$, and $S_{-}$, $S_{0}$ and $S_{+}$).  The iteration estimates of the ground-state energy $\mathcal{E}$ and its first and second $\lambda$ derivatives (at $\lambda=\lambda_0$) are computed by 
\begin{align}
& \oE_{0}  \label{Eav} \\[.3em]
& \frac{\oE_{+}-\oE_{-}}{2\varepsilon}  \label{Ep} \\[.3em]
& \frac{\oE_{+}-2\oE_{0}+\oE_{-}}{\varepsilon ^{2}}  \label{Epp}
\end{align}
respectively, where the averaging (implied by the bar) is done in the manner of (\ref{aver}).  Note that each of the three averages uses its own set of weights (slightly different from the other two), namely $\exp(-\tau S_{-})$, $\exp (-\tau S_{0})$, and $\exp (-\tau S_{+})$; from now on, we denote these $W_{-}$, $W_{0}$, and $W_{+}$ respectively.
\item After thousands of iterations, we convert each of these into the corresponding \emph{grand-mean} estimate (by ordinary averaging).  Repeating with several different values of $\tau$ and extrapolating to $\tau=0$ yields the ultimate estimates (and their standard errors) of the three quantities.
\end{enumerate}

An important issue to resolve is whether the three nearly parallel paths of any such `triplet' of configurations will remain at $\varepsilon$ -- proportional distances throughout the simulation, or whether their separation can slowly but steadily increase.  Luckily (and perhaps somehow surprisingly), it appears that this process is rather stable and the probability of any of the three companions diverging from the other two is extremely small.  But it \emph{can} happen, and when it does (even to a single triplet of an ensemble of thousands), the damage is large and permanent (the tight bond between originally parallel configurations will never be restored -- they now go their own separate ways).

To minimize the probability of this happening, and to expediently fix it when it does happen, we propose to reduce the distance from the `leading' configuration (the one with $\lambda = \lambda_{0}$, say $\xx_{0}$) to each of its two companions (say $\xx_{+}$ and $\xx_{-}$) by a factor similar to (\ref{Efac}); that means that, at the end of every
iteration (and for each triplet of the ensemble). $\xx_{+}$ is replaced by
\begin{equation}
\xx_{0} + (\xx_{+} - \xx_{0}) \cdot \left( 1-\dfrac{a^{\prime }~\tau }{1-\ln \tau }\right)  \label{stabil}
\end{equation}
and similarly for $\xx_{-}$.This not only alleviates the problem, but also reduces (quite significantly) the statistical error of all the estimates, at the cost of introducing yet another$\tau$-proportional bias (which is then automatically removed by extrapolating to $\tau =0$). Since this also effectively reduces the value of$\varepsilon$ in (\ref{Ep}) and \eqref{Epp} by the factor of
\begin{equation}
r:=1-\dfrac{a^{\prime }~\tau }{1-\ln \tau }
\end{equation}
one should adjust the two formulas accordingly; it helps reducing the corresponding $\tau$-related bias.

The $a^{\prime }$ constant needs to be adjusted empirically; making it too small leaves us with substantial statistical error, but make it too large and the $\tau$ -related bias thus introduced becomes impossible to remove by extrapolation.\bigskip

\noindent\textbf{Note}: To find corresponding estimates of the $3^{\text{rd}}$ and $4^{\text{th}}$ derivative, we have to add two extra parallel companions, one with $\lambda =\lambda _{0}-2\varepsilon$ and the other with $\lambda=\lambda _{0}+2\varepsilon$ (let us call the corresponding iteration means of the local energy $\oE_{--}$ and $\oE_{++}$ respectively).  The iteration estimates of the two derivatives are then (incorporating the $r$ factor)
\begin{equation}
\frac{\oE_{++}-2\oE_{+}+2\oE_{-}-\oE_{-}}{2\varepsilon^{3}r^{3}}
\end{equation}
and
\begin{equation}
\frac{\oE_{++}-4\oE_{+}+6\oE_{0}-4\oE_{-}+\oE_{--}}{%
\varepsilon ^{4}r^{4}}
\end{equation}
respectively; the rest of the procedure remains the same (only $\varepsilon$ must then be kept, more conservatively, in the $10^{-3}$ to $10^{-2}$ range).

\subsection{Example (harmonic oscillator)}

In our first `toy' example (taken from \cite{vrbik}) we consider a one-dimensional quantum harmonic oscillator of a unit mass and the spring constant given by a (rather arbitrarily chosen) function of $\lambda$, namely
\begin{equation}
\frac{1+\lambda }{1-\lambda }
\end{equation}
Analytically solving the corresponding eigenvalue equation%
\begin{equation}
-\frac{1}{2}\frac{\dd^{2}}{\dd x^{2}}\Phi (x)+\frac{x^{2}}{2}\frac{1+\lambda }{1-\lambda }\Phi(x)=\mathcal{E}\,\Phi(x)
\end{equation}
for the lowest eigenvalue yields
\begin{equation}
\mathcal{E}=\frac{1}{2}\sqrt{\frac{1+\lambda }{1-\lambda }}  \label{exact}
\end{equation}
where
\begin{equation}
\Phi(x) \varpropto \exp (-\mathcal{E}\,x^{2})
\end{equation}
implying that, at $\lambda=0$, the ground-state energy and its first four $\lambda$-derivatives are equal to $0.5$, $0.5$, $0.5$, $1.5$, and $4.5$ respectively.

Using the trial function
\begin{equation}
\Psi(x)=\left(1+\frac{2}{5}\frac{1+\lambda }{1-\lambda }x^{4}\right)^{-1}
\end{equation}
the variational (and thus visibly inaccurate) solution for the ground-state energy becomes $\sqrt{10}/3\simeq 1.0541$ times larger than its exact value (\ref{exact}), with the same relative increase in the$\lambda$-derivatives.
The drift and local-energy functions are easily computed to be
\begin{align}
F(x) &= \frac{8(1+\lambda )x^{2}}{5(1-\lambda )+2x^{4}(1+\lambda )} \\[.5em]
E(x) &= \frac{(1+\lambda )x^{2}\left( 145(1-\lambda )^{2}-60(1-\lambda^{2})x^{4}+4(1+\lambda )^{2}x^{8}\right) }{2(1-\lambda )\left(5(1-\lambda )+2(1+\lambda )x^{4}\right) ^{2}}
\end{align}
In the actual simulation, these need to be evaluated at $\lambda=-2\varepsilon$, $-\varepsilon$, $0$, $\varepsilon$, and $2\varepsilon$, where $\varepsilon=0.03\sqrt{\tau}$.

\begin{figure}[htbp]
\centering
\includegraphics[width=.7\linewidth]{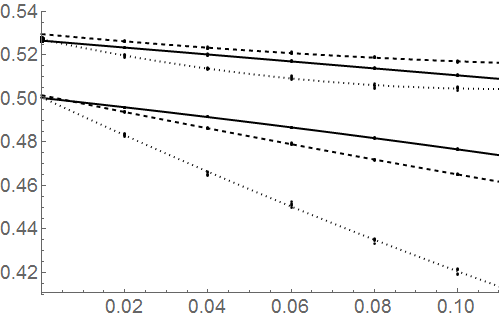}
\caption{Variational and exact estimation of $\mathcal{E}$, $\dfrac{\dd\mathcal{E}}{\dd \lambda }$, and $\dfrac{\dd^{2}\mathcal{E}}{\dd \lambda ^{2}}$.}
\label{Fig1}
\end{figure}

\begin{figure}[htbp]
\centering
\includegraphics[width=.7\linewidth]{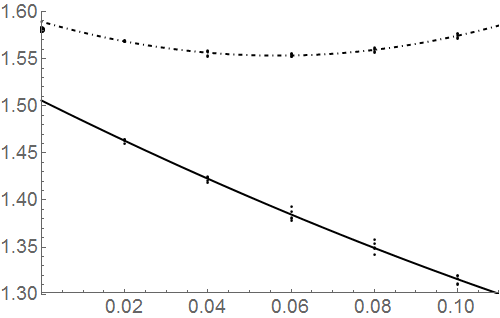}
\caption{Variational and exact estimation of $\dfrac{\dd^{3}\mathcal{E}}{\dd\lambda ^{3}}$}
\label{Fig2}
\end{figure}

\begin{figure}[htbp]
\centering
\includegraphics[width=.7\linewidth]{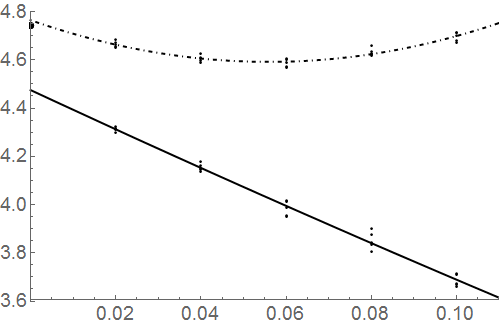}
\caption{Variational and exact estimation of $\dfrac{\dd^{4}\mathcal{E}}{\dd\lambda ^{4}}$}
\label{Fig3}
\end{figure}

Using $10,000$ configurations, five values of $\tau$, namely $0.10$, $0.08$, $0.06$, $0.04$, and $0.02$ (it is advantageous to carry out the simulation in this order, to expedite equilibration), $40,000$ iterations at each $\tau$ (these have been divided into $5$ blocks of $8,000$, yielding $5$ nearly independent estimates at each $\tau$), $a=1.5$ in (\ref{Efac}), $b=0.03$ in (\ref{epsi}) and $a=1$ in (\ref{stabil}), the last transformation being applied to each of the \emph{four} companions of the leading configuration, we get results of Figure \ref{Fig1} (estimating the ground-state energy and its first two $\lambda$ derivatives -- the solid, dashed and dotted line respectively), Figure \ref{Fig2} (the third derivative) and Figure \ref{Fig3} (the fourth one).
The agreement with true values is fairly good.

For comparison, we also give (in the same three graphs, using dashed lines in Figure \ref{Fig2} and \ref{Fig3}) the corresponding results of the variational simulation (which is achieved by simply replacing all weighed averages of the type (\ref{aver}) by \emph{regular} averaging; the $S(\xx)$ quantities are then no longer needed).  The target values (computed analytically) of the variational extrapolation have been indicated by half-disks attached to the vertical scale.

\section{Dealing with singularities}

When simulating molecules, the problem we usually encounter is that both the local energy and the drift function possess several singularities; the local energy becomes infinite (with either sign, depending on the trial function) when an electron approaches a nucleus or when two electrons closely approach each other.  It is always possible to remove these singularities by a careful adjustment of trial-function parameters, but a more expedient solution is to simply truncate the values of $E(\xx)$ to stay between%
\begin{align}
&E_{\min }-\frac{c}{\sqrt{\tau }} \notag \\
&E_{\max }+\frac{c}{\sqrt{\tau }}  \label{trunc}
\end{align}%
where $E_{\min}$ and $E_{\max }$ are the smallest and largest values (respectively) of the local energy function \emph{after} excluding the \emph{outliers} (these can be easily established from the corresponding empirical
histogram), and $c$ is suitably chosen constant.  To be more specific, this means that any value of $E(\xx)$ which is bigger than the upper limit of the (\ref{trunc}) interval is replaced by this upper limit, and similarly for the lower limit (whenever $E(\xx)$ is too small).  These truncated $E(\xx)$ values are then used in the computation of $S(\xx)$
and the corresponding $W(\xx)$ weights.

There are also the (inevitable, due to Pauli exclusion principle) singularities of $\FF(\xx)$ at the \textsc{nodes} (surfaces in the $\xx$ space where $\Phi(\xx)=0$) of the trial function; this is a problem shared by all trial functions and can be easily dealt with by truncating all components of $\FF$ in a manner of (\ref{trunc}), but with its own limits.

Neither of these modifications affects the final results (each of them only introduces yet another $\tau$-related bias which then disappears with the $\tau \rightarrow 0$ extrapolation).

A similar issue arises when computing \eqref{Ep} and \eqref{Epp}; the corresponding outliers can be huge, thus ruining the resulting accuracy.  To be able to deal with these, one first needs to replace the \eqref{Ep} and \eqref{Epp} estimators by
\begin{equation}
\overline{\frac{E_{+}-E_{-}}{2\varepsilon }}-\tau \cdot \overline{\left(E_{0}-\oE_{0}\right) \cdot \frac{S_{+}-S_{-}}{2\varepsilon }}
\label{firs}
\end{equation}
and
\begin{align}
&\overline{\frac{E_{+}-2E_{0}+E_{-}}{\varepsilon ^{2}}}-2\tau \cdot \overline{\left( \frac{E_{+}-E_{-}}{2\varepsilon }-\overline{\frac{E_{+}-E_{-}}{2\varepsilon }}\right) \cdot \frac{S_{+}-S_{-}}{2\varepsilon }} \label{seco} \\
&-\tau \cdot \overline{\left( E_{0}-\oE_{0}\right) \cdot \frac{S_{+}-2S_{0}+S_{-}}{\varepsilon ^{2}}}+\tau ^{2}\cdot \overline{\left( E_{0}-\oE_{0}\right) \cdot \left( \frac{S_{+}-S_{-}}{2\varepsilon }-\overline{\frac{S_{+}-S_{-}}{2\varepsilon }}\right) ^{2}}  \notag
\end{align}
respectively, where now the bars imply averaging with respect to the $W_{0}$ weights \emph{only}.  This enables us to further truncate, individually (i.e. before any averaging), the values of
\begin{equation}
\frac{E_{+}-E_{-}}{2\varepsilon },\ \frac{E_{+}-2E_{0}+E_{-}}{\varepsilon^{2}},\ \tau \cdot \frac{S_{+}-S_{-}}{2\varepsilon },~\text{and}~\tau \cdot \frac{S_{+}-2S_{0}+S_{-}}{\varepsilon ^{2}}
\end{equation}
again in the manner of (\ref{trunc}).  This appears to solve the problem, introducing yet another $\tau$-related and thus inconsequential bias.

The corresponding formulas for the third and higher (and also mixed, when having several parameters) derivatives can be found in \cite{vrbik}.  To understand where these come from, just differentiate (\ref{aver}) with respect to $\lambda$ once to get (\ref{firs}), twice to get (\ref{seco}) etc., keeping in mind that both $E(\xx^{\prime })$ and $S(\xx^{\prime })$ are functions of $\lambda$. This results in a non-trivial combination of quantities called \textsc{cumulant}s; it is the price to pay for interchanging the two operation (ensemble averaging and numerical differentiation).  Utilizing cumulants in this manner appears to be yet another unique aspect of our technique.

Finally (as already mentioned), there is now the danger that one (or more) of the ensemble's triplets approaches a node so closely that one of its three companions crosses the node and the corresponding drift then sends it in \emph{opposite} direction to the other two (as $\FF$ always points \emph{away} from a node); this means that the three (up to that point `parallel') paths will suddenly start separating from each other.  Even though this is now partially alleviated by treating the triplet as an outlier (since it inevitably becomes one) and truncating its contribution to our averages, we clearly cannot afford to keep on creating more of these indefinitely.  The solution has already been provided by bringing the three companions of \emph{all} triplets closer together by the (\ref{stabil}) transformation -- in addition to keeping the simulation more stable in general, this also restores `infinitesimal' separation of any `broken' triplet in a handful of iterations.

\subsection{LiH example}

In this example, we estimate the first \emph{two} derivatives of ground-state energy with respect to $\lambda$, which is defined as the length of the distance between the two nuclei of the \textsf{LiH} molecule (namely \textsf{Li} and \textsf{H}, having atomic numbers of $3$ and $1$ respectively), initialized at $\lambda =3$ (using atomic units).  The molecule has $4$ electrons, two with spin up and two with spin down; the trial function we use here has been taken from \cite{vrbik} and is very simple: a product of two Slater determinants built out of four atomic orbitals, further multiplied by a product of four Jastrow factors (for pairs of electrons with opposite spin).  This trial function yields $-8.029$ as the variational estimate of the ground-state energy (whose exact value is $-8.070$); using \eqref{aver} instead of regular averaging changes this estimate to $-8.062$. The reason why we are still above the correct value is
due to our trial function having slightly incorrect nodes, as mentioned earlier.

\begin{figure}[htbp]
\centering
\includegraphics[width=.7\linewidth]{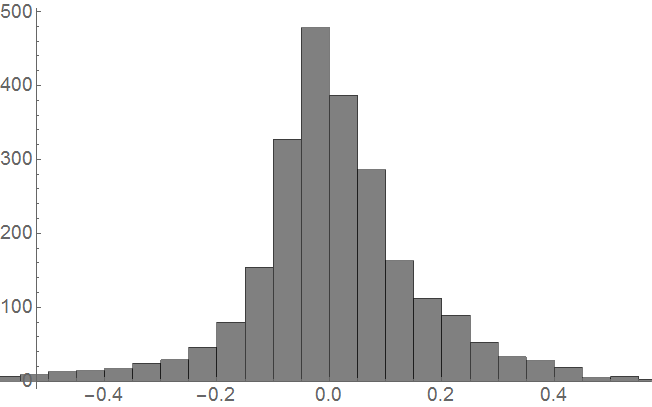}
\caption{Histogram of $\dfrac{E_{+}-E_{-}}{2\varepsilon}$ values.}
\label{Fig4}
\end{figure}

For the actual simulation we used the same set of $\tau$ values and the same $\varepsilon$ as in the harmonic-oscillator example, but this time the number of configurations was $2500$, and so was the number of iterations at each $\tau$ (divided into $5$ blocks of $500$).  Also, we had to use a new value of $a=1/2$, $b=0.03$, and $a^{\prime }=1/3$ (extrapolating to $\tau =0$ becomes too difficult with higher values of $a^{\prime }$; smaller values lead to increasing statistical error) and the following truncations (we list the target quantity with the corresponding lower and upper limit)
\begin{equation}
\begin{tabular}{|c|c|c|}
\hline
Target & Lower & Upper \\
\hline
\hline
$E(\xx)$ & $-9.2-\dfrac{0.1}{\sqrt{\tau }}$ & $-7+\dfrac{0.1}{\sqrt{\tau }}$ \\[.65em]
$\dfrac{E_{+}-E_{-}}{2\varepsilon }$ &$-0.4-\dfrac{0.01}{\sqrt{\tau }}$ & $0.4+\dfrac{0.01}{\sqrt{\tau }}$ \\[.65em]
$\dfrac{E_{+}-2E_{0}+E_{-}}{\varepsilon ^{2}}$ &$-0.5-\dfrac{0.01}{\sqrt{\tau }}$ &$0.5+\dfrac{0.01}{\sqrt{\tau }}$ \\[.65em]
$\tau \cdot \dfrac{S_{+}-S_{-}}{2\varepsilon }$ &$-1-\dfrac{0.02}{\sqrt{\tau }}$ &$1+\dfrac{0.02}{\sqrt{\tau }}$ \\[.65em]
$\tau \cdot \dfrac{S_{+}-2S_{0}+S_{-}}{\varepsilon ^{2}}$ &$-2-\dfrac{0.04}{\sqrt{\tau }}$ &$2+\dfrac{0.04}{\sqrt{\tau }}$ \\[.65em]
\hline 
\end{tabular}
\end{equation}
Furthermore, each component of every $\FF(\xx)$ was truncated to stay within the 
\begin{equation}
\left( -3-\dfrac{0.1}{\sqrt{\tau }},\, 3+\dfrac{0.1}{\sqrt{\tau }}\right)
\end{equation}
interval.  All these limits have been selected based on the corresponding histogram; as an example, Figure \ref{Fig4} displays an ensemble's worth of $(E_{+}-E_{-})/(2\varepsilon)$ values (save a handful of outliers).

The results are shown in Figure \ref{Fig5} (the first derivative, which should be close to zero) and Figure \ref{Fig6} (the second derivative, called the harmonic constant; its true value is about $0.0666$ atomic units); numerically, our two estimates are $-0.0006\pm 0.0006$ and $0.0676\pm 0.0013$, in good agreement with empirical values (the slightly negative first derivative implies that our $\lambda =3$ needs to be corrected to $3+0.0006/0.0676=3.009$, in a reasonable agreement with the observed value of $3.015$).
\begin{figure}[htbp]
\centering
\includegraphics[width=.7\linewidth]{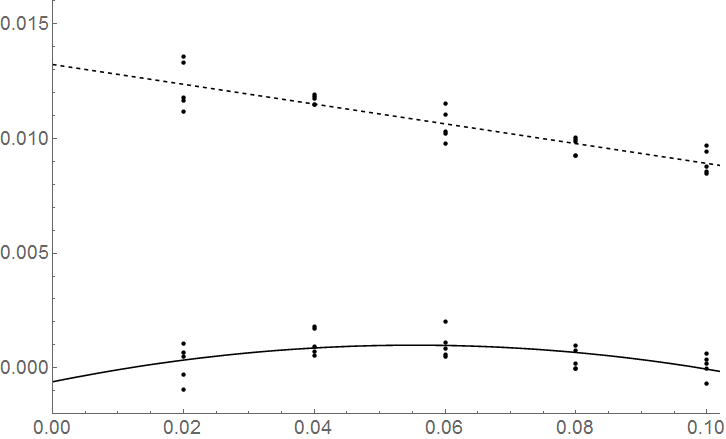}
\caption{Variational and exact estimation of $\dfrac{\dd E}{\dd\lambda }$}
\label{Fig5}
\end{figure}

For comparison, in the same two graphs we also show (dashed lines) the corresponding variational results, computed based on
\begin{equation}
\overline{\frac{E_{+}-E_{-}}{2\varepsilon r}}
\end{equation}
(a single-iteration estimate of the first derivative) and
\begin{equation}
\overline{\frac{E_{+}-2E_{0}+E_{-}}{(\varepsilon r)^{2}}}
\end{equation}
(of the second one), where the bar now indicates \emph{regular} averaging. The resulting estimates are (not surprisingly) rather inferior.
\begin{figure}[htbp]
\centering
\includegraphics[width=.7\linewidth]{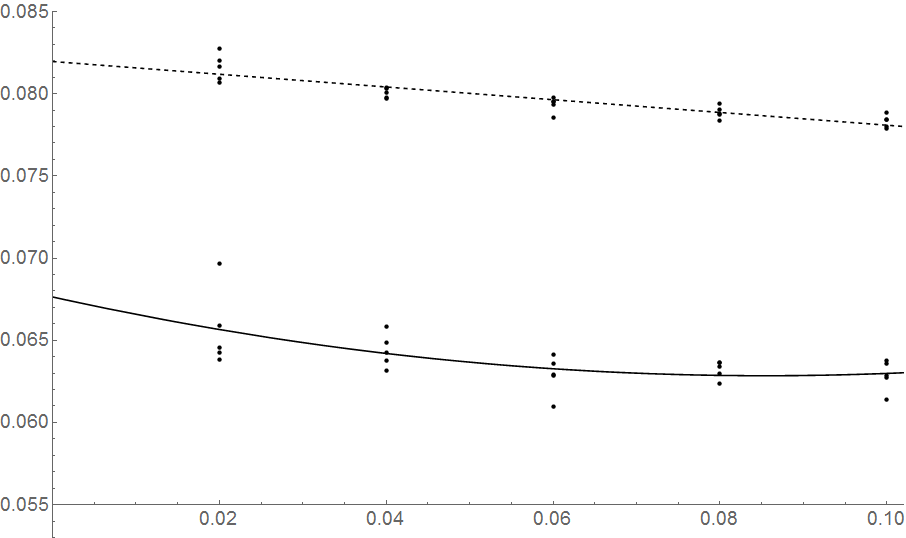}
\caption{ Variational and exact estimation of $\dfrac{\dd^{2}E}{\dd\lambda ^{2}}$}
\label{Fig5}
\end{figure}

\section{Conclusion}

We have demonstrated a new technique of \emph{parallel} sampling (distinct from \emph{correlated} sampling of existing literature) for computing derivatives of ground-state energy with respect to a given parameter using a guiding function which is allowed to vary with the parameter.  This is of particular importance when establishing a molecule's geometry and corresponding vibrational properties.  The results are subject only to the fixed-node approximation and the inevitable statistical error.  We thus claim that, in principle, one can achieve practically exact results by improving
the guiding function and increasing the length of computation.  This is in contrast to our previous work \cite{vrbik} which, to estimate the same geometry-related properties, had to introduce an extra approximation of unknown bias.

\end{document}